\documentstyle[10pt,epsfig]{article}
\begin{document}
\textwidth 6.75in
\textheight 8.5in

\begin {center}
{\Large Decays of $\sigma$, $\kappa$, $a_0(980)$ and $f_0(980)$}

\vskip 5mm
{D.V.~Bugg\footnote{email address: D.Bugg@rl.ac.uk}},   \\
{Queen Mary, University of London, London E1\,4NS, UK}
\end {center}
\vskip 2.5mm

\begin{abstract}
Ratios of coupling constants for these decays are
compared with $q\bar q$ predictions and Jaffe's $q^2\bar q^2$
model.
In both models, the predicted ratio $g^2(\kappa \to K\pi)/
g^2(\sigma \to \pi \pi )$ is much too small.
Also, for $q\bar q$, the predicted ratio
$g^2(\kappa \to K\eta ')/g^2(\kappa \to K\pi)$
is much larger than observed.
Both models fail for $g^2(f_0 \to KK)/g^2(a_0 \to KK)$.
This ratio requires that $f_0$ has a dominant $KK$ component.
%For $\sigma$ and $\kappa$, three independent results all favour
%Jaffe's model quite strongly.
It arises naturally because the $f_0$ pole lies very
close to the $KK$ threshold, giving its  wave function a long $KK$
tail.

\vspace{5mm}

\end{abstract}

\section {Introduction}
There are conflicting opinions whether $\sigma$, $\kappa$,
$a_0(980)$ and $f_0(980)$ are predominantly molecular states,
$q\bar q$ or 4-quark.
There are now extensive data for their
coupling constants to pseudoscalars:
(i) for $\sigma$ and $f_0$ to $\pi \pi$, $\eta \eta$ and $KK$,
(ii) for $\kappa$ to $K\pi$, $K\eta$ and $K\eta '$,
and (iii) for $a_0(980)$ to $\eta \pi$ and $KK$.
The objective here is to compare all ratios of coupling constants
with predictions for $q\bar q$ and $q^2\bar q^2$ states.

The $\sigma$ pole has been known to generations of theorists, who
extracted it from data on $\pi \pi$ elastic scattering, see the
summary given by Markushin and Locher [1].
The E791 group then observed it as a peak in $D^+ \to \pi ^+\pi ^-
\pi ^+$ [2].
Higher statistics data from BES for $J/\Psi \to \omega \pi ^+ \pi ^-$
now provide a better determination of the pole position
$M - i\Gamma /2 = (541 \pm 39) -i(252 \pm 42)$ MeV [3].
If it is a $q\bar q$ state, one would expect a brother with $I=1$
at a similar mass, whereas the $a_0(980)$ is over 400  MeV
heavier.

Jaffe proposed that $\sigma$ and its relatives
are $q^2\bar q^2$ states [4].
His suggestion is that there is a pairing interaction forming
S-wave diquarks in the flavour 3 configuration: $ud$, $ds$ and $us$.
Then 3 and $\bar 3$ make a colourless nonet.
The $\sigma$ is the $I = 0$ member $u\bar d d\bar u$,
the $\kappa ^+$ is $u\bar s d \bar d$, $a_0(980)$ is
$s \bar s (u\bar u -d\bar d)\sqrt {2}$
and $f_0(980)$ is $s\bar s (u\bar u + d\bar d)/\sqrt {2}$.
This scheme neatly explains why $a_0$ and $f_0$ are nearly degenerate in
mass and heavier than the $\sigma$ by twice the mass of the $s$-quark.
It also fits in neatly with the intermediate mass of the
$\kappa$. From the latest  combined analysis of E791, BES and LASS
data, the $\kappa $ pole is at $(750^{+30}_{-55}) -i(342 \pm 60)$ MeV
[5].

There is support for Jaffe's scheme from recent Lattice QCD
calculations of Okiharu et al. [6].
They find configurations at large radii consisting of a $qq$
pair joined by a flux tube to a $\bar q \bar q$ pair.
At small radii, they find two meson-meson pairs.
The implication is that the  massive ``dressed" $q^2\bar q^2$
configuration can decay by fission to two lighter pions at small
separations.

Fig. 1(a) shows diagrams for decays of $q\bar q$ states.
Fig. 1(b) shows fall-apart decays for four-quark states.
In principle, four-quark combinations can make not only
nonets but higher SU(3) multiplets, which Jaffe discusses in detail.
One can view his hypothesis as a final-state interaction which favours
the nonet configuration.

%Fig. 1
\begin {figure}  [h]
\begin {center}
\centerline{\hspace{0.5cm}\epsfig{file=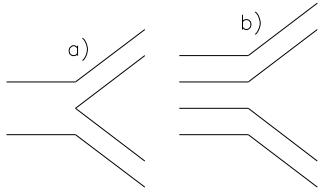,width=7cm}}
\vskip -15mm
\caption{Decays of (a) $q\bar q$, (b) $q^2\bar q^2$ states.}
\end {center}
\end {figure}

An alternative possibility is that $\sigma$ and its relatives are
`molecular' states created by long-range meson exchanges. There is a
long history of proposals along these lines [7-10].
If one takes the $K$-matrix element in the
$s$-channel from these Born terms, the unitarised amplitude $K/(1 -
i\rho K)$ reproduces the observed $\pi \pi$ S-wave quite well up to 1
GeV and beyond. In all cases, the attraction from exchanges is barely
sufficient to produce resonances. Indeed, for $a_0(980)$, the J\" ulich
group, Janssen et al. [10] find only a virtual state. Likewise,
unpublished calculations by Zou and myself find that there is not quite
enough attraction from $K^*(890)$ and $\rho (770)$ exchange to generate
$f_0(980)$ or $a_0(980)$. Because of this, widths predicted for
$a_0$ and $f_0$ are particularly sensitive to meson coupling constants.
As a result, no comparison will be made here with the `molecular'
hypothesis.

Although the amplitudes for $\pi \pi$ and $K\pi$ elastic scattering can
be predicted adequately from meson exchanges, the nonet of
$\rho$, $\omega$, $K^*(890)$ and $\phi$ cannot be
predicted in this way.
Instead they appear as CDD poles [11].
This suggests that $\sigma$, $\kappa$, $a_0(980)$ and $f_0(980)$ are
not regular $q\bar q$ states, although their formation may be related
to short-range $q\bar q$ components, as in the approach of van
Beveren et al [12-14]. A new comparison with their model is in
preparation.

Section 2 introduces some caveats.
Both $f_0(980)$ and $a_0(980)$ lie close enough to the $KK$ threshold
that they must contain substantial $KK$ components, resembling the
long-range tail of the deuteron.
It will be shown that
the ratio of these long-range components is close to 2, and can
account for the experimental ratio $g^2(f_0 \to KK)/g^2(a_0 \to KK)$.
The experimental data also suggest mixing between $\sigma$ and
$f_0(980)$ in a mass range centred on the $KK$ threshold.

Section 3 shows that both $q\bar q$ and $q^2\bar q^2$
schemes fail to account for several ratios of $g^2$.
Both schemes predict
$g^2(f_0(980) \to KK)/g^2(a_0(980) \to KK)$ close to 1, in
disagreement with the experimental value $2.15 \pm 0.4$.
In Jaffe's scheme, this problem may be remedied by taking
$f_0(980)$ to be dominantly $KK$.
There is, however, a residual problem in describing
$\sigma \to KK$.
The $q\bar q$ hypothesis fails to fit the ratio $g^2(f_0(980) \to KK)/
g^2(a_0(980) \to KK)$ even when $f_0(980)$ is taken to be
dominantly $KK$.
Both schemes fail to account for the branching ratio
$g^2(\kappa \to K\pi )/g^2(\sigma \to \pi \pi )$.

Section 4 points out the possible existence of a narrow
state at the $\eta ' \eta '$ threshold.
This does not fit into Jaffe's model.
The summary in section 5 attempts to reach some conclusions.

From this point onwards, $a_0(980)$ and $f_0(980)$ will be
abbreviated to $a_0$ and $f_0$ unless there is possible confusion
with other states.

\section {$f_0(980)$ must have a large $KK$ component}
At a mass just below the $KK$ threshold, both $f_0$ and
$a_0$ must have a long range tail due to small binding
energy.
T\" ornqvist [15] discusses this issue.
His eqn. (15) gives a formula for the $KK$ component in the wave
function:
%Eqn.
\begin {equation}
\psi = \frac {|q\bar q> + \sum _i [-(d /d s)\mathrm{Re}~\Pi
_i(s)]^{1/2}|A_iB_i>}
         {1 - \sum _i (d /d s) \mathrm{Re} ~ \Pi _i(s)},
\end {equation}
where $AB$ stands for molecular components $KK$,
$\eta \eta$, $\pi \eta$, etc.
The quantity $\Pi$ is the propagator of the resonance and
$\mathrm{Re}~\Pi _{KK}(s) = g^2_{K\bar K}\sqrt {4m^2_K/s -
1}$ for $s < 4m^2_K$; there is a corresponding term for $\eta \eta$.
[T\" ornqvist's equation is written in terms of $q\bar q$, but could
equally well be reformulated in terms of 4-quark states].
His formula is easily evaluated to find the $K\bar K$ components in
$a_0$ and $f_0$ as functions of $s$.
At the $K\bar K$ threshold, the binding energy goes to
zero and the $KK$ wave function extends to infinity, so the $KK$
fraction $\to 1$.
Using BES parameters [16],
the $f_0(980)$ has a second sheet pole at $(998 \pm 4)
-i(17 \pm 4)$ MeV, very close to the $KK$ threshold; there is a distant
third sheet pole at $(851 \pm 28) - i(418 \pm 72)$ MeV.
The dominance of the narrow second sheet pole is used by Baru et al.
[17] to argue that $f_0(980)$ is mostly a $KK$ bound state pole.
The $a_0(980)$ with parameters derived in Ref. [18] has a second sheet
pole at $1032 -i85$ MeV and a third sheet pole at $968-i245$ MeV.
This is closer to a conventional resonance and further from
the $KK$ threshold.

Results from T\" ornqvist's formula are shown in  Fig. 2 by the dotted
curves.
This figure also shows line-shapes as the full curves.
If the $KK$ component behaves as an inert cloud for radii $> 0.6$ fm,
the mean $KK$ fraction integrated over the line-shape is
$\ge 70\%$ for $f_0$ and $\sim 35\%$ for $a_0$.

%Fig. 2.
\begin{figure} [htb]
\begin{center}
\epsfig{file=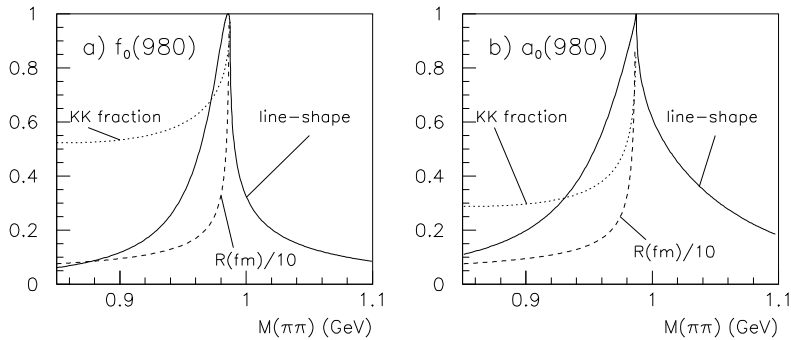,width=12cm}\
\vskip -8mm
\caption{Line shapes of $f_0(980)$ and $a_0(980)$, full curves; the $KK$
fraction in the wave function, dotted curves; $0.1 \times $RMS
radius, dashed curves.}
\end{center}
\end{figure}

\subsection {Further caveats}
It is first necessary to explain the view adopted here for the
broad $\sigma$.
It is a very curious resonance with unusual features.
Achasov and Shestakov were amongst the first to clarify the
relation with chiral symmetry within the framework of the
linear $\sigma$ model [19]. They pointed out that it cannot
be fitted adequately by a simple Breit-Wigner resonance
with a large constant width.
It has a pole at $\sim 540$ MeV, but the observed $\pi \pi$
elastic phase shift goes through $90^\circ$ at $\sim 1 $ GeV.
How are these two facts reconciled?

The clue is that the width is strongly $s$-dependent, with a zero
at the Adler point $s \simeq 0.5m^2_\pi$, just below
threshold.
Experiments on $\pi \pi$ elastic scattering are done at
real values of $s$.
In finding the pole, it is necessary to extrapolate the measured
amplitude off the real $s$-axis.
The Breit-Wigner amplitude fitted to the data has a width of
the form $\Gamma = A(s)(s - s_A)\rho _{\pi \pi }(s)$,
where $\rho$ is the usual Lorentz invariant phase space
$\sqrt {1 - 4m^2_\pi /s}$;
$A(s)$ is a slowly varying function of $s$.
The phase shift goes to 0 at the $\pi \pi$ threshold.
However, the pole lies at $s_0 = 0.23 - i0.27$ GeV$^2$.
In the extrapolation to the pole, the factor $(s - s_A)\rho _{\pi \pi
(s)}$ develops a large phase rotation of $\sim 55^\circ$ near the pole.
Oller drew attention to this earlier [20].
The result is that the pole is approximately $55^\circ$ ahead of the
phase of the amplitude along the real axis; there is a further small
phase variation arising from the slowly varying function $A(s)$, but it
is only a few degrees in practice.
Qualitatively, the broad $\pi \pi$ amplitude measured on the real
$s$-axis may be viewed as a long tail of the pole buried deep in the
complex $s$-plane.
Its phase reaches  $90^\circ$ for real $s \sim 1$ GeV$^2$.
In production data, one sees a peak in the $\pi \pi$ intensity at
$\sim 500$ MeV, but that peak is hidden in $\pi \pi$ elastic scattering
by the Alder zero in that process [21].
Fig. 2(c) of that reference shows a graph of the mass at which the
phase passes $90^\circ$ for complex $s$.

The broad component of the $\pi \pi$ S-wave continues through the
mass range 1 to 2 GeV.
It stretches the imagination to interpret it as the tail of the
$\sigma$ pole.
It may therefore have a further origin in that mass range.
For example, Anisovich et al. [22] argue that is should be
interpreted there as a broad glueball.

For the $\kappa$, the situation is even more extreme.
The phase rotation between the pole and the real $s$-axis is
$\sim 85^\circ$. The long tail of this pole is fitted to
$K\pi$ elastic phase shifts determined in the LASS experiment of
Aston et al. [23].
Because of the very large phase rotation, these phase shifts do not
quite reach $90^\circ$ over the mass range where they have been
measured.

The view being examined here is a narrow one, that the amplitude
for the so-called $\sigma$ in the vicinity of the $KK$ and $\eta \eta$
thresholds may be expressed in terms of just two orthogonal states
$f_0(980)$ and $\sigma$.
Likewise, the amplitude for the $\kappa$
may be expressed in terms of just two orthogonal states
$\kappa$ and $K_0(1430)$.
This could be an over-simplification.
The objective is to see where this view leads.

Resonances have a finite spatial extent.
The $\pi \pi$ S-wave amplitude is known up to 1.9 GeV.
One can take the Fourier transform of the observed $s$-dependence
to determine the radius of interaction.
The result is a rather small RMS radius of 0.45 fm [21].
This is quite enough to produce a large form factor between
the mass of the $\sigma$ pole and 1 GeV, and likewise between the
mass of the $\kappa$ and the $K\eta $ and $K\eta '$ thresholds.
One should remain alert to the fact that coupling parameters
are likely to be $s$-dependent.
It is therefore not realistic to use ratios like
$g^2(a_0 \to \eta \pi)/g^2(\sigma \to \pi \pi)$ for
quantitative purposes, because the poles are too far apart.

Data on $\phi$ radiative decays are analysed in an accompanying paper
[24] and provide a precise measurement of the ratio $g^2(\sigma \to
KK)/g^2(\sigma \to \pi \pi)$ from interference with $f_0(980)$.
This will be taken as a reliable number at the $KK$ threshold.
Data on $\pi \pi \to KK$ are also analysed in Ref. [24].
These data are fitted over a range of masses up to 1.9 GeV;
they appear to confirm
the result from KLOE data within a somewhat larger error.

There is a further caveat which is rarely discussed.
For an isolated $\sigma$ produced without $f_0(980)$, there is a
multiple scattering series for $\sigma \to \pi \pi$ and $KK$.
The $f_0(980) $ has its own multiple scattering series.
In $\pi \pi$ elastic scattering, both the broad $\sigma$ and $f_0(980)$
appear strongly at 1 GeV in Cern-Munich data [25].
The multiple scattering series then contains additional terms of the
form $\sigma \to \pi \pi \to f_0$ and vice versa.
These cross-terms are likely to lead to dynamical mixing
of $f_0(980)$ and $\sigma$ unless the overlap of their wave function
happens to be zero.
However, this mixing can vary from process to process, depending on
how much of each is produced in the formation reaction.
For example, in $J/\Psi \to \omega \pi \pi$, the $\sigma $ is produced
strongly, but there is little or no $f_0(980)$ [3].
In $J/\Psi \to \phi \pi ^+ \pi ^-$, the $f_0(980)$ is produced
strongly with a small $\sigma$ amplitude accompanying it [16].

In elastic scattering, both $\sigma$ and $f_0(980)$ are produced
strongly.
The analysis of data in the accompanying paper [24] shows
that a substantial $\sigma \to KK$ component is needed near
the $KK$ threshold with $g^2(\sigma \to KK)/g^2(\sigma \to \pi \pi ) =
0.6 \pm 0.1$.
However, it appears to be somewhat localised near the $KK$ threshold.
KLOE data on $\phi \to \gamma \pi ^0 \pi ^0$
will not tolerate a $\sigma \to KK$ component with the large
width of the $\sigma$;
this amplitude must be attenuated strongly below $\sim 800$ MeV.
It may be fitted using a rather strong form factor $\exp -\alpha
|k|^2$ where $k$ is KK centre-of-mass momentum.

Above the $KK$ threshold, data on $\pi \pi \to KK$ again require a
rather strong form factor to fit the observed strong decrease in the
cross section from 1 to 1.8 GeV.
The result is a broad peak in the coupling to $KK$ over a mass range
roughly 800 to 1300 MeV.
A straightforward possibility is that there is mixing between
$f_0(980)$ and $\sigma$, peaking there.
For this reason, the analysis in the next Section will focus on ratios
of $g^2$ only close to $KK$ and $\eta \eta$ thresholds.

\subsection {Systematic errors for coupling constants}
Flatt\' e formulae have been used in fitting $a_0$ and $f_0$,
but ignoring coupling of $a_0(980)$ to $\eta '\pi$ and $f_0(980)$
to $\eta \eta$.
Tests adding these couplings suggest that effects are small
compared with errors assigned by the experimental groups.
The BES data have been refitted allowing the $\eta \eta$
coupling explicitly, but the data suggest no coupling to this
channel.
The main source of systematic error for the $f_0$ is the effect of
possible mixing with $\sigma$.
This mixing depends on unknown wave functions.
The coupling constant for $\sigma \to KK$ is particularly
sensitive to this mixing, but errors assigned in Ref. [24]
are intended to cover the range of possible form factors
fitted to both $\sigma$ and $f_0$.

The $\sigma$ pole was determined by the BES collaboration [3]
fitting several different Breit-Wigner forms. The quoted
systematic errors cover this range of possibilities.
Some authors have raised the possibility of unknown `non-resonant
backgrounds' in the $\sigma$.
However, without educated guesses about such possible backgrounds,
there is no limit to the possible range of parameters which can
be fitted.
The approach adopted here is to fit a simple empirical $s$-dependent
width including the Adler zero; errors for the fitted parameters cover
the likely range of possibilities.
The same approach is adopted for the $\kappa$.
There, the main problem in fitting parameters is
unknown mixing with $K_0(1430)$.
However, the latest refit to LASS, BES and E791 data [5] arrives at
a consistent picture from the three sets of data; the range of
parameters fitting all three sets of data will be used to cover
possible systematic errors. There is a small systematic discrepancy
with LASS data around 1.2 GeV, and its effect on possible coupling
to $K\eta$ will be discussed in Section 3.3.

Parameters of $\sigma$ and $\kappa$ are completely insensitive to
precise locations of the Adler zeros.

\section {Comparison of $g^2$ with $q\bar q$ and Jaffe's model}
It is necessary first to discuss the selection of data used for
this comparison.
The $\sigma$ pole will be taken from the high statistics data of
BES, where there is a clearly visible peak with a well defined
mass and width [3].
The $\kappa$ pole will be taken from the combined analysis of
E791 and BES production data and LASS phase shifts [5].
The $f_0(980)$ appears as a strong peak in BES data for
$J/\Psi \to \phi \pi ^+\pi ^-$ and $\phi K^+K^-$ [16].
Its width is precisely determined by the $\pi \pi$ peak
because of the very good mass resolution.
The signal is clearly visible near threshold in the $KK$ channel.
The ratio of events in this peak to that in the $\pi \pi$ peak
determines the branching ratio $g^2(KK)/g^2(\pi \pi)$ accurately.

The $a_0(980)$ is  subjected to detailed scrutiny in an accompanying
paper [24] which compares a fit to Kloe data on $\phi \to \gamma \eta
\pi ^0 $ with an earlier determination from Crystal Barrel data [18].
There is agreement within errors, and values of $g^2$ are taken from
the combined analysis.
Most other experimental determinations quoted by the Particle Data
group are fitted to a Breit-Wigner resonance of constant
width, an assumption far from reality.

The comparison of $g^2$ made here is motivated by a similar
comparison for well known $q\bar q $ states such as
$\rho (770)$ and $K^*(890)$.
After allowing for effects due to identical particles (discussed
in detail in the next section), the prediction for
$g^2(K^*(890) \to K\pi)/g^2(\rho \to \pi \pi )$ is 3/4.
This agrees well with experiment if one allows a P-wave
Blatt-Weisskopf centrifugal barrier for both decays with a reasonable
radius of 0.5 fm.
For $\sigma$ and its relatives, no centrifugal barrier is involved
between mesons, so a comparison of $g^2$ should be a meaningful test
of the models.

\subsection {Formulae}
Formulae for coupling of $\sigma$, $\kappa$, $a_0$ and
$f_0$ to $q\bar q$ have been given by Anisovich, Anisovich and
Sarantsev [22].
Corresponding formulae for $q^2 \bar q^2$ are given by Jaffe in
Table 7 of his publication [4].
However there is an important subtlety concerned with these
formulae for identical particles $\pi \pi$ and $\eta \eta$.

Consider the Breit-Wigner amplitude $a$ for a process involving
non-identical particles: $K^+K^- \to a_0 \to \eta \pi$,
\begin {equation}
a = \frac {g_{K^+K^-}g_{\eta \pi }} {m^2 - s - i(g^2_{\eta
\pi } \rho _{\eta \pi} + g^2_{KK}\rho _{KK})}.
\end {equation}
The  integrated cross section involves an integral $4\pi$ over the solid
angle.
For $K\bar K \to f_0 \to \pi ^0 \pi ^0$, isospin Clebsch-Gordan
coefficients combining two isospins 1 to $I=0$ lead to a final state
$(\pi ^+\pi ^- - \pi ^0 \pi ^0 + \pi ^-\pi ^+)/\sqrt {3}$.
The $\pi ^+ \pi ^-$ cross section may be determined by counting
$\pi ^+$ over the whole solid angle. At a particular angle $\theta$,
there are two amplitudes $\pi ^+(\theta)\pi ^-(\theta +\pi)$ and
$\pi^- (\pi + \theta)\pi ^+(\theta)$ which add coherently.
The integrated intensity over angles is $(4/3)4\pi$.
For $\pi ^0 \pi ^0$, there are again contributions
$\pi ^0 (\theta ) \pi ^0 (\pi + \theta ) $ and $\pi ^0 (\pi + \theta )
\pi ^0 (\theta ) $, but the angular integration should now be done over
only one hemisphere, to avoid counting both $\pi ^0$
from a single event.
The result is $(4/3)2\pi$.
The total $\pi \pi$ integral is $(6/3)4\pi$.
So the identity of the pions leads to a doubling of the $\pi \pi$
branching ratio and $g^2_{\pi \pi}$, and likewise for $\eta \eta$.

The hypothesis to be tested here is that all members of a $q\bar q$
nonet have the same coupling constant $g$
(apart from effects of identical particles and Clebsch-Gordan
coefficients).
Anisovich and Sarantsev include the factor 2 for identical particles
explicitly into branching ratios for $\pi \pi$ and $\eta \eta$.
That convention will be followed here.
However, Jaffe gives formulae for amplitudes without the identity
factor and leaves the user to put it in.

For $q\bar q$ states, the following linear combinations will be
used:
\begin {eqnarray}
\sigma &=& n\bar n \cos \phi + s\bar s \sin \phi \\
f_0 &=&   -n\bar n \sin \phi + s\bar s \cos \phi ,
\end {eqnarray}
where $n\bar n = (u\bar u + d\bar d)/\sqrt {2}$.
Diagrams of Fig. 1(a) for decay of $I=0$ $q\bar q$ states lead to
a final state
\begin {eqnarray}
\nonumber
A =
  &[&u(\bar uu + \bar dd + \sqrt {\lambda} s\bar s)\bar u
+ d(\bar uu + \bar dd + \sqrt {\lambda} s\bar s)\bar d]\frac {\cos
\phi}{\sqrt {2}} \\
  &+&s[\bar uu + \bar dd + \sqrt {\lambda} s\bar s]\sin \phi .
 \end {eqnarray}
The factor $\sqrt {\lambda }$ is introduced by Anisovich and
Sarantsev to allow for possible differences between $\bar nn$
and $\bar ss$.
There is a one-to-one correspondence between each term in
this series and the possible diagrams of Fig. 1(a).

The $\eta$ and $\eta '$ will be written as
\begin {eqnarray}
\eta &=& n\bar n \cos \theta _P - s\bar s \sin \theta _P \\
\eta '&=& n\bar n \sin \theta _P + s\bar s \cos \theta _P \\
\eta _0 &=& \eta \cos \theta _P +\eta '\sin \theta _P \\
\eta _s &=& -\eta \sin \theta _P +\eta '\cos \theta _P ,
\end {eqnarray}
where $\theta _P$ is the pseudoscalar mixing angle; the value
$\sin \theta _P = 0.608 \pm 0.025$ will be used [26].
A straightforward expansion of eqn. (5) gives
\begin {eqnarray}
\nonumber
A = &[&\eta \eta \cos ^2 \theta _P + \pi ^0 \pi ^0 - \pi ^- \pi ^+ - \pi
^-\pi ^+ + \sqrt {\lambda}(K^0\bar K^0 - K^-K^+)]\frac {\cos
\phi}{\sqrt {2}} \\ &+&[K^0\bar K^0 - K^-K^+ +\sqrt {\lambda }\eta
\eta \sin ^2\theta _P] \sin \phi ;
\end {eqnarray}
contributions from
$\eta \eta '$ and $\eta \eta$ have been omitted for simplicity.

The $a_0^+$ leads to a final state
\begin {equation}
F(a_0) = u(\bar uu + \bar dd + \sqrt {\lambda }\bar ss)\bar d;
\end {equation}
after using G parity to eliminate $\pi \pi$ final states
(or alternatively the Pauli principle), the surviving amplitude is
\begin {equation}
F(a_0) = \frac {1}{\sqrt {2}}(\eta _0\pi ^+ + \pi ^+\eta _0) +
\sqrt {\lambda }K^+\bar K^0.
\end {equation}
For the $\kappa ^+$,
\begin {eqnarray}
F(\kappa ^+) &=& u(\bar uu + \bar dd +
\sqrt {\lambda }\bar ss )\bar s \\
&\to & \frac {1}{ \sqrt {2}}(\eta _0 - \pi ^0)K^+ + \pi ^+K^0
+ \sqrt {\lambda }\eta _sK^+.
\end {eqnarray}
Resulting $q\bar q$ branching ratios (integrated
over charge states) are shown in column 2 of Table 1.

\begin{table}[htb]
\begin {center}
\begin{tabular}{ccc}
\hline
Ratio of $g^2$& $q\bar q$ & $q^2\bar q^2$  \\\hline
$(\kappa \to K\pi)/(\sigma \to \pi \pi )$
&$ 1/(2\cos^2 \phi )$ & $1/(2\cos ^2 \phi)$ \\
$(\kappa \to K\eta)/(\kappa \to K \pi )$
& $(c-\sqrt {2{\lambda }}s)^2/3 $  & $c^2/3$ \\
$(\kappa \to K\eta ')/(\kappa \to K \pi )$
&$(s+\sqrt {2\lambda }c)^2/3$ & $s^2/3$ \\
$(a_0 \to \pi \eta )/(a_0 \to KK)$
& $2c^2$ &$s^2$ \\
$(a_0 \to \pi \eta ')/(a_0 \to KK)$
& $2s^2$ &$c^2$ \\
$(\sigma \to \eta \eta)/(\sigma \to \pi \pi )$
& $(c^2 + \sqrt {2\lambda }s^2\tan \phi )^2$/3 &
$(c^2 - \sqrt {2}cs\tan \phi )^2/3$ \\
$(\sigma \to KK)/(\sigma \to \pi \pi )$
& $(\sqrt {\lambda }
+ \sqrt {2}\tan \phi )^2/3$ & $(1/3)\tan ^2\phi $ \\
$(f_0 \to \eta \eta)/(f_0 \to \pi \pi )$
& $(c^2 - \sqrt {2\lambda }s^2\cot \phi )^2/3$
& $(c^2 + \sqrt {2}cs\cot \phi )^2/3$ \\
$(f_0 \to KK)/(f_0 \to \pi \pi )$
& $(\sqrt {\lambda }
-  \sqrt {2}\cot \phi )^2/3$ & $(1/3)\cot ^2\phi  $ \\
$(f_0 \to KK)/(a_0 \to KK)$
&  $(\sin \phi - \sqrt {2/ {\lambda }}
\cos \phi)^2$ & $\cos ^2 \phi$ \\\hline
\end{tabular}
\caption{Ratios of $g^2$ predicted by $q\bar q$ and $q^2\bar q^2$
models; $c = \cos \theta _P$, $s = \sin \theta _P$.}
\end {center}
\end{table}

Jaffe's  model requires a non-strange $I=0$ component $N$ which
may be written
\begin {eqnarray}
N &=&
(1/2) (u\bar d d\bar u + d\bar u d\bar d + u\bar u d\bar d
+ d\bar d u\bar u)\\
&=& (1/2)(\eta _0\eta _0 + \pi ^0 \pi ^0 -\pi ^-\pi ^+ -
\pi ^+\pi ^-).
\end {eqnarray}
There is an orthogonal state with hidden strangeness
\begin {eqnarray}
S &=&
(1/2) (u\bar u s\bar s + d\bar d s\bar s + u\bar s s\bar u
+ d\bar s s\bar d)\\
&=& (1/2)(\sqrt {2}\eta _0\eta_s - K^+K^- + K^0\bar K^0).
\end {eqnarray}
Further states are
\begin {eqnarray}
a_0^+ &=&
(1/\sqrt {2}) (u\bar d s\bar s + u\bar s s\bar d)\\
&=& (1/\sqrt {2})(\pi ^+\eta _s  + K^+\bar K^0), \\
\kappa ^+ &=& (1/\sqrt {2})(u\bar s d\bar d + d\bar s u\bar d) \\
&=& (1/2)[K^+(\eta_0 + \pi ^0) + \sqrt {2}K^0\pi ^+].
\end {eqnarray}
The quantities $N$ and $S$ replace $n\bar n$
and $s\bar s$ in eqns. (3) and (4).
The third column of Table 1 shows branching ratios for Jaffe's
model.

\subsection {Conclusions from $f_0(980)$, $a_0(980)$ and $\sigma$}
\begin{table}[htb]
\begin {center}
\begin{tabular}{cccc}
\hline
 Ratio of $g^2$ & $q\bar q$ & $q^2\bar q^2$ & Expt
\\\hline
$(f_0 \to \eta \eta )/(f_0 \to \pi \pi )$ & $0.37 \pm 0.14
 \, {\rm or} \, 0.83 \pm 0.09$ & $3.11 \pm 0.08 \,{\rm or}\, 1.07
 \pm 0.18$ & $< 0.33$  \\
$(f_0 \to KK )/(a_0 \to KK)$  & $1.11 \pm 0.04 \, {\rm or}
\, 2.96 \pm 0.03$ & $0.93 \pm 0.01$ & $2.15 \pm 0.4$  \\
$(\sigma \to KK)/(\sigma \to \pi \pi )$& $0.69 \pm 0.02 \,
 {\rm or} \, 0.02 \pm 0.01$ & $0.03 \pm 0.01$ & $0.6 \pm 0.1$ \\
$(\sigma \to \eta \eta)/(\sigma \to \pi \pi )$& $0.21 \pm 0.01 \,
 {\rm or} \, 0.04 \pm 0.01$ & $0.06 \pm 0.02 \, {\rm or} \,
 0.23 \pm 0.02$ & $0.20
 \pm 0.04$  \\\hline
\end{tabular}
\caption{Ratios of $g^2$ for $q\bar q$ and $q^2\bar q^2$
models predicted from $g^2(f_0 \to KK)/g^2(f_0 \to \pi \pi)$, compared
with experimental values from Ref. [24];
alternative solutions are with $\phi$ positive (first solution)
or negative (second).}
\end {center}
\end{table}

The parameter $\lambda$ of Anisovich and Sarantsev was preserved in
Table 1 for reference purposes;
however, it does not systematically improve agreement
with experiment, so it will be set to 1.
The initial objective is to show that the data for these decays
are inconsistent with either $q\bar q$ or $q^2\bar q^2$ for
$f_0(980)$.

Let us start from the ratio
$g^2(f_0 \to KK)/g^2(f_0 \to \pi \pi) = 4.21 \pm 0.46$, which is well
determined from recent BES data on $J/\Psi \to \phi \pi ^+\pi ^-$ and
$\phi K^+K^-$ [16];
statistical and systematic errors have been combined in quadrature.
These data lead to two possibilities for the mixing angle $\phi$.
For $q\bar q$, they are $(17.3 \pm 0.7)^\circ$ or $(-29.0 \pm
2.0)^\circ$.
The errors cover purely experimental errors for the ratio
$g(f_0 \to KK)/g^2(f_0 \to \pi \pi)$;
in Table 2, errors from this source are propagated and added in
quadrature with errors from  the pseudoscalar mixing angle $\theta _P$.
The first solution, $\phi = +17.2^\circ$ agrees with experiment for
three ratios, but fails for $g^2(f_0 \to KK)/g^2(a_0 \to KK)$.
The second solution, $\phi = -29.0^\circ$ fails for three ratios.

The $q^2\bar q^2$ scheme leads to two solutions with
$\phi = \pm (15.7 \pm 0.9)^\circ$.
Neither solution agrees with all experimental ratios.
The branching ratio of $f_0 \to \eta \eta$ is far above the
experimental limit and the branching ratio for $\sigma \to KK$ is
far below experiment.

In view of the prediction from Section 2 that $f_0(980) $ should
contain a large $KK$ component, we immediately turn to the case
where $f_0(980)$ is pure $KK$:
\begin {eqnarray}
\sigma &=& n\bar n \cos \phi + KK \sin \phi \\
f_0 &=&   -n\bar n \sin \phi + KK \cos \phi \\
S &=& KK.
\end {eqnarray}
Results are shown in Table 3.
For $q\bar q$, the only change to eqn. (10) is the disappearance of
the term $\eta \eta \sin ^2 \theta _P$.
There is therefore no change to values of $\phi$, and entries 2 and 3
remain unchanged.
Entry 1 is marginally improved and entry 4 is slightly worse.
There is no improvement in the ratio
$g^2(f_0 \to KK)/g^2(a_0 \to KK)$.

 \begin{table}[htb]
\begin {center} \begin{tabular}{cccc} \hline
 Ratio of $g^2$ & $q\bar q$ & $q^2\bar q^2$ & Experiment
\\\hline
$(f_0 \to \eta \eta )/(f_0 \to \pi \pi )$ & $0.13 \pm 0.01$ & $0.13
 \pm 0.01$ & $< 0.33$  \\
$(f_0 \to KK )/(a_0 \to KK)$  & $1.11 \pm 0.04 \, {\rm or} \,
2.96 \pm 0.03$ & $1.73 \pm 0.03$ & $2.15 \pm 0.4$  \\
$(\sigma \to KK)/(\sigma \to \pi \pi )$& $0.69 \pm 0.02 \,
 {\rm or} \, 0.02 \pm 0.01$ & $0.02 \pm 0.02$ & $0.6 \pm 0.1$ \\
$(\sigma \to \eta \eta)/(\sigma \to \pi \pi )$& $0.13 \pm 0.01$
& $0.13 \pm 0.01$ & $0.20 \pm 0.04$ \\
\hline
\end{tabular}
\caption{Ratios of $g^2$
for $q\bar q$ and $q^2\bar q^2$ models predicted with $f_0$ pure $KK$.}
\end {center}
\end{table}

For Jaffe's model, $S$ of eqn. (18) is replaced by
$(1/\sqrt {2})(K^0\bar K^0 - K^+K^-)$.
The value of $\phi$ changes to $\pm (21.7 \pm 1.2)^\circ$;
there is an improvement in the ratio
$g^2(f_0 \to KK)/g^2(a_0 \to KK)$ to a value within one
standard devation of experiment.
However, entry 3 is still far from experiment.

One can try to improve the agreement for the $q^2\bar q^2$
scenario by including a small amount of the hidden
strange component $S$ of Jaffe's model, using
\begin {eqnarray}
\sigma
&=& N\cos \phi + \frac {\alpha S +KK} {\sqrt {1 + \alpha ^2}}\sin \phi \\
f_0 &=& -N \sin \phi +\frac {\alpha S + KK}{\sqrt {1 + \alpha ^2}} \cos
\phi .
\end {eqnarray}
However, it turns out that there is no solution which gives agreement
with both $f_0 \to \pi \pi$ and
$\sigma  \to KK$.
The best that can be achieved is to increase $g^2(\sigma \to KK)/
g^2(\sigma \to \pi \pi)$ to 0.29, still a factor 2 smaller than
experiment.

\subsection {Results for the $\kappa$}

\begin{table}[htb]
\begin {center}
\begin{tabular}{cccc}
\hline
Ratio of $g^2$& $q\bar q$ & $q^2\bar q^2$ & Expt  \\\hline
$(\kappa \to K\pi)/(\sigma \to \pi \pi )$ & 0.55  & 0.58
& $2.14 \pm 0.28$ \,{\rm to}\, $1.35 \pm 0.10$ [3,5]\\
$(\kappa \to K\eta)/(\kappa \to K \pi )$  & $0.004 \pm 0.005$ &
$0.20 \pm 0.01$  & $0.06 \pm  0.02$ [5] \\
$(\kappa \to K\eta ')/(\kappa \to K \pi )$& $1.00 \pm 0.01$
& $0.13 \pm 0.01$ & $0.29 \pm 0.29$ [5] \\
$(a_0 \to \pi \eta )/(a_0 \to KK)$        & $1.21 \pm 0.06$
& $0.40 \pm 0.03$ & $0.75 \pm 0.11$ [25] \\
$(a_0 \to \pi \eta ')/(a_0 \to KK)$        & $0.79 \pm 0.06$
& $0.60 \pm 0.03$ & - \\\hline
\end{tabular}
\caption{Ratios of $g^2$ predicted by $q\bar q$ and $q^2\bar q^2$ models
and experimental values.}
\end {center}
\end{table}

Table 4 shows predictions
for $\sigma$, $\kappa$ and $a_0$.
For the first entry, the predicted
ratio is almost the same for $q\bar q$ and $q^2\bar q^2$;
the best values of $\phi$ are chosen from Table 3.  At the
position of the $\kappa$ pole, $|\rho _{K\pi }| = 0.821$ and at the
$\sigma$ pole $|\rho _{\pi \pi }| = 0.936$. Using the width observed
for the $\sigma$ pole by BES, $504 \pm  84$ MeV, both $q\bar q$ and
$q^2\bar q^2$ predict a $\kappa$ width of $236 \pm 39$ MeV. Such a
narrow $\kappa$ would be extremely conspicuous; it is completely ruled
out by the data, which require a width roughly a factor 3 larger [5].
Nonetheless, the experimental ratio $g^2(\kappa \to K\pi)/g^2(\sigma
\to \pi \pi)$ quoted in Table 4 requires some explanation.
The first value $2.14 \pm 0.28$ is obtained from the conventional
expression $M\Gamma /|\rho |$ at the pole.
However, it is debatable what effect the Adler zero has on
the width.
Experimentally, the width is parametrised as $A(s) (s - s_A)$,
where $A(s)$ is a slowly varying exponential factor preventing
the width from increasing continuously with $s$.
The $\sigma$ pole lies closer to its Adler zero than the
$\kappa$ pole.
The Adler zero might therefore suppress the width.
An extreme view  is to factor the term $(s - s_A)$ out of
the width and examine the ratio of $A(s)$ at the pole.
This gives the second result of Table 2, $1.35 \pm 0.10$.
Incidentally, the small error arises from a cancellation
between correlations involved in finding the pole position.

Consider next decays $\kappa \to K\eta '$.
For $q\bar q$, entry 3 of Table 4 shows that the predicted ratio
$g^2(\kappa \to K\eta ') /g^2(\kappa \to K\pi)$ is large and $\sim 2.5$
standard deviations away from experiment.
It is possible that the $K\pi$ signal may be attenuated by a form
factor, but this would make the disagreement worse. If the
broad $\kappa$ signal were to couple strongly to $K\eta '$, one should
see a strong dispersive effect in the vicinity of this threshold. There
is no sign of any such effect in the data.
It therefore appears that there is a discrepancy with the $q\bar q$
hypothesis.

Entry 2 shows predictions for $K\eta$.
Both are small. Neither model makes an accurate prediction, though
they both give small numbers like experiment.
The $q^2\bar q^2$ model does not fare so well.
However, a warning is that Ref. [5] points out that the fit to LASS data
is not perfect around 1.2 GeV, fairly close to the $K\eta$
threshold.
This problem may well arise because the $s$-dependence being
fitted presently to the data is the simplest possible and
may be over-simplified; the error quoted for the experimental value is
purely statistical and does not allow for possible systematic error.
The simple fact is that there is no evidence for structure in the broad
$\kappa \to K\pi$ at the $K\eta$ threshold.
It would be valuable to have data directly for the $K\eta$ channel.

The fourth entry of Table 5 compares $g^2(a_0 \to \eta \pi)$ with
$g^2(a_0 \to KK)$.
The experimental ratio is rather well known from the combination
of Crystal Barrel data and KLOE data [24].
It is over three standard devations larger than predicted by Jaffe's
model. The $q\bar q$ prediction is higher than experiment. However, the
momentum available in $\eta \pi$ decays is 325 MeV/c and a form factor
with an RMS radius of 0.75 fm could bring the $q\bar q$ prediction into
agreement with experiment; for the $q^2\bar q^2$ case, such a form
factor would make matters worse.
From Section 2, the $a_0(980)$ must contain a $KK$ component of
$\sim 35 \%$.
This is neither small nor large.
It is possible that a more refined model allowing for this $35\%$ KK
component might change the level of agreement with the $q^2\bar q^2$
hypothesis, but it is not presently clear how to construct such a
model.
The present conclusion is that $q\bar q$ gives
better agreement with experiment.

Table 4 shows predictions for $a_0 \to \pi \eta '$.
Presently there are no data for this ratio.
Such data are important to complete the picture.

\subsection {Discussion}
Neither $q\bar q$ nor Jaffe's model gives reasonable
agreement with experiment.
The failure to predict the ratio $g^2(f_0 \to KK)/g^2(a_0 \to KK)$
may reasonably be attributed to the fact that $f_0$ has a dominant
$KK$ cloud.
With this correction, Jaffe's model predicts a ratio within
$1\sigma$ of experiment.
However, $q\bar q$ still fails to predict this ratio.
This is because eqn. (12) predicts $g^2(a_0 \to KK)$ a factor 2
larger than eqn. (20) of Jaffe's model.

However, the critical point where both $q\bar q$ and Jaffe's model
fail seriously
is the prediction $g^2(\kappa \to K\pi)/g^2(\sigma \to
\pi \pi ) = 1/(2\cos ^2 \phi).$
This is in complete contradiction with experiment.

An alternative scenario is that $\sigma$, $\kappa$, $a_0$ and
$f_0$ are driven by meson exchanges [7-10].
These calculations show that $\pi \pi$ and $K\pi$ phase shifts
may be reproduced by taking Born terms from the meson exchanges
and unitarising the amplitude using the K-matrix.
The calculations provide a valuable clue.
All these resonances are only just bound.
Coupling constants of mesons need to be adjusted (within their errors)
to reproduce phase shifts for $\pi \pi$ and $K\pi$ and resonance masses
and widths for $f_0$ and $a_0$.
As coupling constants increase, phase shifts vary more rapidly with
$s$, i.e. resonances become {\it narrower}.
This is the reverse of what happens for $q\bar q$ states treated as
CDD poles.
It seems likely that this point is at the root of the disagreement
between data and the comparisons made here with $q\bar q$ and
$q^2\bar q^2$ models.

Oller [27] makes a comparison with a scheme along these lines
where K-matrix elements for $\sigma$ and $\kappa$ are taken
from Chiral Perturbation Theory.
Resonances are then generated dynamically.
This gives a more promising agreement with SU(3) and he
claims to obtain reasonable agreement with a $q\bar q$ nonet.
However, he predicts a coupling of $f_0 \to \eta \eta$ (which is
nearly the same as his $\eta _8 \eta _8$) almost as large as to $KK$.
The new BES data for $f_0(980)$ rule out that possibility, which would
lead to a dramatic fall in the $f_0 \to KK$ and $\pi \pi$ signals at
the $\eta \eta $ threshold.

\section {Structure at the $\eta \eta '$ threshold?}
An important experimental question is whether there is a further
$s\bar s s \bar s$ state.
This is foreign to Jaffe's nonet. GAMS have reported tentative
evidence for a narrow state in $\eta \eta '$ at 1914 MeV, almost
exactly the $\eta '\eta '$ threshold [28].
They claim $\Gamma (\pi ^0 \pi ^0 )/\Gamma (\eta \eta ') < 0.1$.
Such a state should decay easily to $\eta \eta '$. If
it were narrow, as GAMS claim ($\Gamma = 90 ^{+35}_{-50}$ MeV ), its
decays to $\eta '\eta '$ would be suppressed by phase space.
This is a similar situation to $f_0(980)$, which appears as a narrow
cusp in $\pi \pi$, but is much more difficult to observe in $KK$,
despite strong coupling to that channel.
Barberis et al. [29] report an $\eta \eta '$ enhancement with
$M = 1934 \pm 16$ MeV, $\Gamma = 141 \pm 41$ MeV, but favour quantum
numbers $J^{PC} = 2^{++}$.
This could well be a different well-known resonance $f_2(1920)$,
seen prominently in decays to $\pi \pi$, $\omega \omega$
and $\eta \pi \pi$.
They also see a threshold enhancement in $\eta ' \eta '$ with
$M = 2007 \pm 24$ MeV, $\Gamma = 90 \pm 43$ MeV; this
could be $f_2(1920)$ ($q\bar q$ $^3P_0$) or its well established
$^3F_2$ partner $f_2(2001)$, observed in all of $\pi \pi$, $\eta \eta$,
$\eta \eta '$ and $f_2\eta$ with consistent mass and width [30].
The key to sorting out this situation is to get high statistics data
on $\eta \eta '$ with good mass resolution.

\section {Summary}
There is evidence that $f_0(980)$ must have a substantial
$KK$ cloud, as predicted in Section 2.
The basic pointer to this conclusion is that the ratio
$g^2(f_0 \to KK)/g^2(a_0 \to KK)$ is at least a factor 2 larger
than can be fitted by either $q\bar q$ or $q^2\bar q^2$.

Otherwise, conclusions are negative; but it may be important to know
what does not work.
This negative conclusion should not be surprising for $q\bar q$, in
view of the fact that masses are far from the usual nonet
configurations such as $\omega $, $\rho$, $K^*(890)$ and $\phi$.
 Predictions for relative widths of $\kappa$ and $\sigma$
fail badly for both $q\bar q$ and Jaffe's model. Meson exchange models
fare better in predicting $\pi \pi$ and $K\pi$ phase shifts. These
models predict that $\sigma$, $\kappa$, $f_0$ and $a_0$ are only just
bound. The large decay widths of $\sigma$ and $\kappa $ reflect this
fact: they decay easily to lighter $\pi\pi$ and $K\pi$ systems.
An approach along these lines will be considered in a separate paper.

\vskip 2mm
{\bf Acknowledgements:} I am grateful to Dr. F. Kleefeld,
Prof. G. Rupp, Prof. E. van Beveren, and Dr. C. Hanhart for extensive
discussions about formulae for branching ratios. Also to Prof. R. Jaffe
for delving into 29 year-old notes for detailed derivation of
amplitudes in his model.

\end {document}